\documentclass[twocolumn,pra,aps,showpacs,nofootinbib,superscriptaddress]{revtex4}
\usepackage{xspace,amsmath,amsfonts,amsthm,amssymb,amsbsy,graphicx,color,setspace}

\newcommand{\ket}[1]{\vert#1\rangle}

\newcommand{\e}[0]{\mathrm{e}}

\begin{document}

\title{Testing nonlocality over 12.4 km of underground fiber with universal time-bin qubit analyzers}

\date{\today}
%\author{F\'elix Bussi\`eres}\email{felix.bussieres@unige.ch, GAP-Optique, University of  Geneva}
\author{F\'elix Bussi\`eres\footnote{Current address: GAP-Optique, University of Geneva, \mbox{e-mail: felix.bussieres@unige.ch}}}
\affiliation{Institute for Quantum Information Science and Department of Physics and Astronomy,
University of Calgary, 2500 University Drive NW, Calgary, Alberta, T2N~1N4~Canada}
\affiliation{Laboratoire des fibres optiques,
D\'epartement de g\'enie physique, \'Ecole Polytechnique de Montr\'eal,
C.P.~6079, Succursale Centre-Ville,
Montr\'eal, Qu\'ebec, H3C~3A7~Canada}
\author{Joshua A. Slater}
\affiliation{Institute for Quantum Information Science and Department of Physics and Astronomy,
University of Calgary, 2500 University Drive NW, Calgary, Alberta, T2N~1N4~Canada}
\author{Jeongwan Jin}
\affiliation{Institute for Quantum Information Science and Department of Physics and Astronomy,
University of Calgary, 2500 University Drive NW, Calgary, Alberta, T2N~1N4~Canada}
\author{Nicolas Godbout}
\affiliation{Laboratoire des fibres optiques, D\'epartement de g\'enie physique, \'Ecole Polytechnique de Montr\'eal,
C.P.~6079, Succursale Centre-Ville, Montr\'eal, Qu\'ebec, H3C~3A7~Canada}
\author{Wolfgang Tittel}
\affiliation{Institute for Quantum Information Science and Department of Physics and Astronomy,
University of Calgary, 2500 University Drive NW, Calgary, Alberta, T2N~1N4~Canada}

\begin{abstract}
We experimentally demonstrate that the nonlocal nature of time-bin entangled photonic qubits persists when one or two qubits of the pair are converted to polarization qubits. This is possible by implementing a novel Universal Time-Bin Qubit Analyzer (UTBA), which, for the first time, allows analyzing time-bin qubits in any basis. We reveal the nonlocal nature of the emitted light by violating the Clauser-Horne-Shimony-Holt inequality with measurement bases exploring all the dimensions of the Bloch sphere. Moreover, we conducted experiments where one qubit is transmitted over a 12.4~km underground fiber link and demonstrate the suitability of our scheme for use in a \textit{real-world} setting. The resulting entanglement can also be interpreted as hybrid entanglement between different types of degrees of freedom of two physical systems, which could prove useful in large scale, heterogeneous quantum networks. This work opens new possibilities for testing nonlocality and for implementing new quantum communication protocols with time-bin entanglement.
\end{abstract}
\pacs{03.65.Ud, % Entanglement and quantum nonlocality
42.65.Lm % Parametric down conversion and production of entangled photon
03.67.Hk % Quantum communication
}

\maketitle

Quantum physics allows physical systems to be correlated in a way that is impossible to describe using a local variable model~\cite{Bell64}. In the case of two physical systems, nonlocal correlations can be produced when they are entangled. From a practical point of view, entanglement is also a powerful resource for the field of quantum communication~\cite{TW01,TG07}. Specifically, photonic entanglement is appealing as it can easily be distributed and, when combined with quantum memories~\cite{LST09}, it can be swapped~\cite{ZZHE93} to create a quantum repeater~\cite{BDCZ98} needed to extend the distribution of quantum states over arbitrarily long distances~\cite{SSRG09}. Furthermore, in the context of a large scale, heterogeneous quantum network~\cite{Kimble08}, photonic entanglement could be used to interface different types of quantum links, encodings, wavelengths and quantum memories. Fortunately, as photonic entanglement exists independently of the choice of encoding, it should persist even if these encoding are converted into each other~\cite{ZZ91}. Photonic entanglement between a polarization qubit and a linear momentum qubit has been shown recently~\cite{MQKJZ09}. While this is interesting for the purpose of quantum computing~\cite{VPMMB07,CLZC07}, the linear momentum encoding is badly suited for long distance transmission as it requires phase stability between two spatial modes. To this end, the polarization and time-bin encodings are much better suited~\cite{UTSW07,MRTZ04}.

\begin{figure}[!h]
\includegraphics{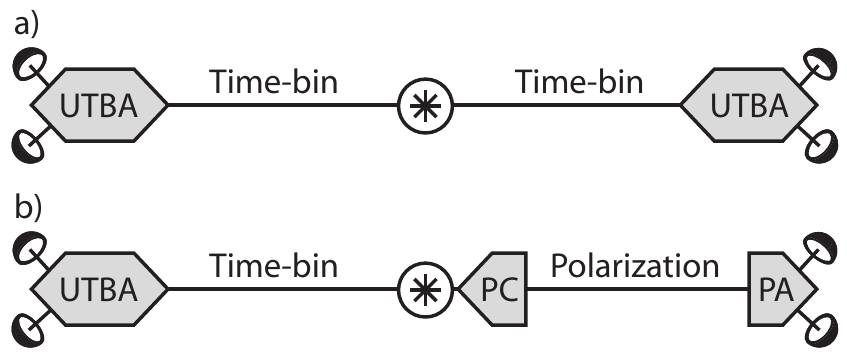}
\caption{a) Source of time-bin entangled qubits, represented by the circled asterisk, with UTBAs, b) A source of hybrid entanglement is obtained by splitting apart one UTBA into a time-bin-to-polarization converter (PC) and a polarization analyzer (PA).} \label{fig:sourcetype} 
\end{figure}

In this article, we experimentally demonstrate that entanglement persists when one or both qubits of a time-bin entangled pair are converted to polarization qubits (for a recently reported, similar investigation see \cite{FTS+10}). Furthermore, this enables the first measurement of time-bin entangled qubits in arbitrary bases, as shown in Fig.~\ref{fig:sourcetype}-a. This qubit conversion is possible by implementing a novel Universal Time-Bin Qubit Analyzer (UTBA), proposed in~\cite{BSBLG06}. If one part of one of the two UTBAs is placed next to the source, as described in Fig.~\ref{fig:sourcetype}-b, this modified source generates hybrid entanglement between a time-bin qubit and polarization qubit. 

We reveal the presence of entanglement through the violation of the Clauser-Horne-Shimony-Holt (CHSH) inequality~\cite{CHSH69} with a variety of measurement settings exploring all the dimensions of the Bloch sphere. To the best of our knowledge, this has never been done before with any encoding. Finally, we demonstrate the persistence of entanglement when one of the qubits travels through an underground 12.4~km fiber link installed between two locations physically separated by 3.3~km. 

The importance of a UTBA is twofold. First, it constitutes a versatile tool to implement quantum communication protocols with time-bin qubits. This is exemplified by our recent implementation of a new loss-tolerant quantum coin-flipping protocol~\cite{BBBG09,BBBGST09}. Second, it should prove useful to experimentally violate a Leggett-type inequality~\cite{Leggett03,Gr07,Branciard07} with time-bin entangled qubits. Indeed, these inequalities require the ability to project qubits onto bases that span all the dimensions of the Bloch sphere in order to rule out nonlocal models than can vio\-late Bell-type inequalities but can not reproduce all the predictions of quantum mechanics. From a more general point of view, this source can also be seen as an interface between heterogeneous quantum communication links, namely free-space and fiber links with possibly different types of encodings and wavelengths, and it should prove useful for quantum networking in general.

We begin by describing the principle of the UTBA, both in its free-space and fiber optics versions. We recall that a time-bin qubit denotes a single photon in superposition of being emitted in two well defined time windows of width $\Delta t$, separated by time~$\tau > \Delta t$, that we shall label \textit{early} $\ket{e}$ and \textit{late} $\ket{\ell}$: $\ket{\psi}  = \cos \theta \ket{e} + \e^{i\phi}\sin \theta \ket{\ell}$~\cite{BGTZ98}. Note that both basis states have identical polarization. In order to project a time-bin qubit onto any basis, one approach is to convert it first to a polarization qubit and then select the basis using wave plates and a polarizing beamsplitter (PBS), as proposed in~\cite{BSBLG06}. 

\begin{figure}
\includegraphics{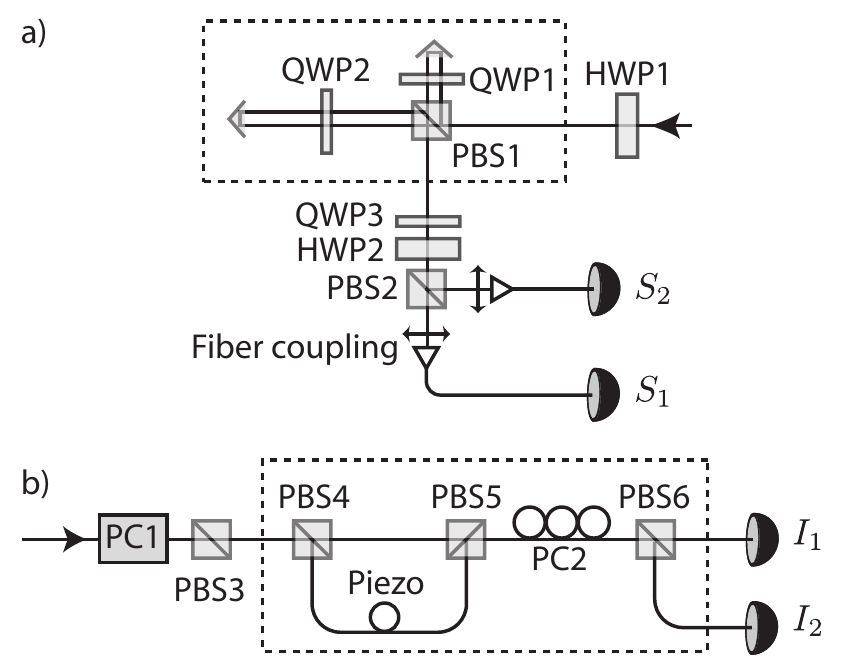}
\caption{a) Free-space UTBA, b) Fiber optics UTBA. The dashed boxes show the sections that were enclosed in a temperature controlled box.} \label{fig:UTBA} 
\end{figure}

Fig.~\ref{fig:UTBA}-a shows the free-space UTBA designed for a wavelength around 810~nm that we built for this experiment. The polarization of the incoming time-bin qubit is first rotated using the half-wave plate HWP1 to $45^{\circ}$ with respect to the linear polarization transmitted by the input polarizing beamsplitter PBS1. Each time-bin qubit component is then separated equally and travels along the short and long arms of a folded Mach-Zehnder interferometer using retroreflectors, featuring a travel time-difference~$\tau = 1.4~\text{ns}$. After being transmitted through the quarter-wave plates QWP1 and QWP2, the light exits through PBS1. At this point, the photon emerges in three chronologically ordered time slots separated by~$\tau$ that we label \textit{early}, \textit{middle} and \textit{late}. In the middle slot, the initial time-bin qubit is mapped onto a polarization qubit, i.e. $\ket{\psi} \rightarrow \cos \theta \ket{V} + \e^{i(\phi + \phi_{A})}\sin \theta \ket{H}$, where $\phi_{A}$ is an additional phase picked-up in the interferometer. This polarization qubit is then analyzed in any basis using the quarter-wave plate QWP3 followed by the half-wave plate HWP2, the polarizing beamsplitter PBS2 and the Si-based single photon detectors $S_{1}$ and $S_{2}$. For increased thermal stability, the optical components of the interferometer are glued on a Zerodur glass plate with a nearly zero linear thermal expansion coefficient. The glass plate is enclosed in a temperature controlled box. To realize the source of hybrid entanglement described in Fig.~\ref{fig:sourcetype}-b, one would physically separate QWP3, HWP2, PBS2 and the detectors from the rest and implement a proper transmission link in between.

\begin{figure*}
\includegraphics{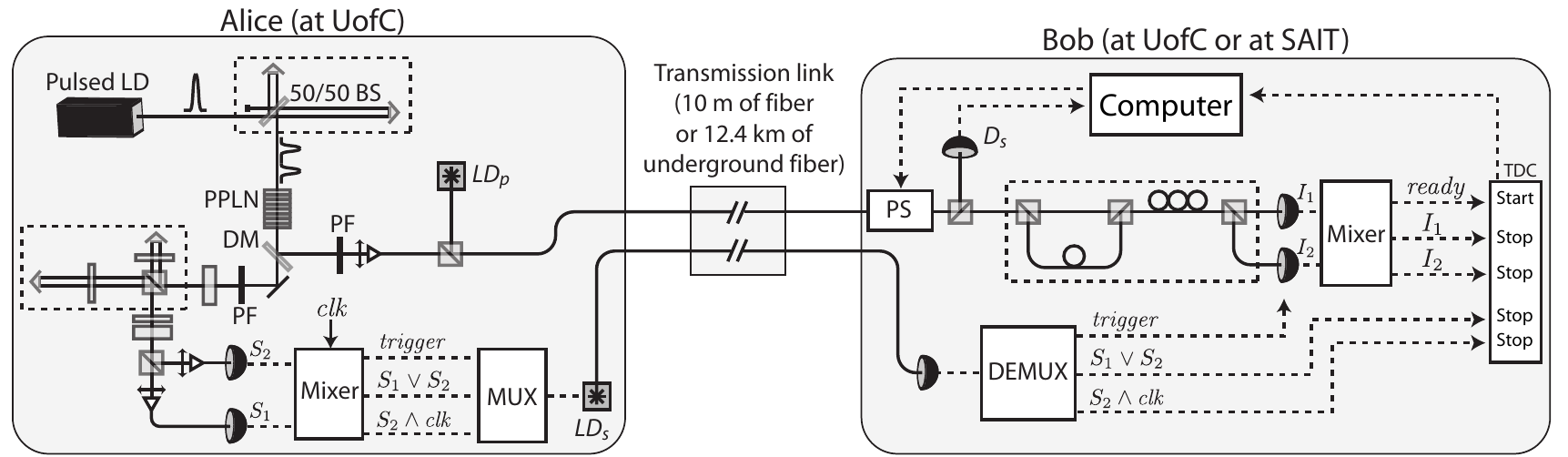}
\caption{A laser diode generates 50~ps pulses at 530.6~nm wavelength and 20~MHz repetition rate that are sent through a folded Mach-Zehnder interferometer with a travel time-difference of 1.4~ns. The pulses emerge in a balanced superposition of two well defined time-bins labelled \textit{early} and \textit{late} and then propagate into a nonlinear, periodically poled lithium niobate crystal (PPLN), thereby creating time-bin entangled qubits at 811.9 and 1531.4~nm wavelengths through spontaneous parametric downconversion. The two qubits are separated at the dichroic mirror (DM) and the pump is filtered out using high-pass filters (PF). The 811.9~nm qubit is measured directly at Alice's using her free-space UTBA. The 1531.4~nm qubit is coupled into the transmission link and sent to Bob, who measures it with his fiber UTBA. The travel time-difference of all three interferometers is the same within a fraction of the coherence time of the photons. Detection events are acquired by a time to digital converter (TDC) that measures delays between a start signal and several stop signals. The detection signals from Alice's free-running Si-based single photon detectors are pre-processed with an electronic mixer. When performing the experiment over the underground fiber link, the signals exiting the mixer are time-multiplexed (MUX) and used to trigger a laser diode $LD_{s}$. The laser pulses are sent over a second fiber link parallel to the one used for the single photon signals. The signals are time-demultiplexed at Bob's (DEMUX) and then used for synchronization and data processing. The \emph{trigger} signal is generated when a detection at either $S_{1}$ or $S_{2}$ occurs and emerges synchronously with the laser clock (\textit{clk}). This signal is used to gate Bob's InGaAs-based single photon detectors during a 7~ns activation window. The \emph{ready} signal, which is emitted only when both detectors are ready to detect, is used to start the TDC\@. This ensures that the statistics are not biased by the dead-time of Bob's detectors. The stop signals $S_2 \wedge \textit{clk}$ and $S_1 \vee S_2$,  where $\wedge$ denotes the logical AND and $\vee$ the logical OR, are used to determine which detector clicked and at what time, respectively. The detections at $I_1$ and $I_2$ also serve as stop signals. This allows us to register all possible coincidence detection events in the middle time slot. The duration of the latter was narrowed down to a time that varied between $0.4$ to $0.8$~ns.
} \label{fig:source} 
\end{figure*}
The fiber optics version of the UTBA, designed for a wavelength around 1530~nm, is shown in Fig.~\ref{fig:UTBA}-b. First, the polarization controller PC1 is used to maximize the transmission through PBS3. The output of PBS3 is a polarization maintaining fiber that is connected to the input port of PBS4 with its slow axis is aligned at $45^{\circ}$ with respect to the transmitted polarization. The outputs of PBS4 couple light into the slow axis of the short and long fibers of the Mach-Zehnder interferometer that also features the travel time-difference~$\tau$. A section of the long arm is wrapped and glued around a circular piezo actuator that is used to control the relative optical phase accumulated during transmission through the two arms of the interferometer. Both paths recombine at PBS5 and are coupled into a single mode fiber. Then, a \textit{Lef\`evre} polarization controller PC2~\cite{Lefevre80} and PBS6 are used to select the measurement basis in the middle time slot, and detection occurs in one of the InGaAs-based single photon detectors $I_{1}$ or $I_{2}$. The interferometer was placed in a temperature controlled box for increased stability. 

The alignment of the paddles of PC2 was performed in the following way. First, a strong pulse of light prepared in the early time-bin was sent into the interferometer. Let us consider the component that travels through the short arm. After PBS4, its polarization is horizontal ($\ket{H}$). Let $\cos \theta \ket{H} + \e^{i\phi} \sin \theta \ket{V}$ be the polarization right before PBS5. The energy of the pulse transmitted through PBS5 can be monitored by replacing $I_{1}$ with a fast classical detector connected to an oscilloscope. By rotating the paddles, the area of the light pulse can be maximized: this indicates that $\theta = 0$. Then, the paddles can be rotated to reduce the area by a factor $\cos^2 \theta$, which allows inferring the value of $\theta$. Note that the value of $\phi$ cannot be measured in this way: this would require performing a complete polarization state analysis. However, as we explain later, our measurements did not require measuring the exact value of $\phi$. 

We note that this UTBA requires the polarization of the time-bin qubit to be set properly at the input. This may seem to defeat the purpose of using the time-bin encoding, which is immune to birefringence in the transmission link~\cite{BGTZ98}.  This is, however, not entirely true as the polarization at the input of the UTBA only requires to be set to horizontal up to an arbitrary phase, whereas polarization encoding requires the horizontal and vertical polarizations' relative phase to be set to zero, which is more challenging to achieve.

Our source of time-bin entangled qubits is detailed in Fig.~\ref{fig:source}. Time-bin entangled photonic qubits in the state
\begin{equation} 
\ket{\psi} = \frac{1}{\sqrt{2}}\left( \ket{e}_{A}\ket{e}_{B} + \ket{\ell}_{A}\ket{\ell}_{B}\right) \label{eqn:time-bin}
\end{equation}
were created, where $\ket{e}_{A(B)}$ and $\ket{\ell}_{A(B)}$ represent the \emph{early} and \emph{late} time-bin qubit states of Alice (Bob). The wavelength and FWHM bandwidth of Alice's qubit was measured using a monochromator and found to be $811.9 \pm 1.6$~nm. From energy conservation, and assuming a monochromatic pump, Bob's qubit is calculated to be encoded at $1531.4 \pm 5.6$~nm wavelength.

We performed two set of experiments. In the first one, Alice and Bob were placed side by side in our laboratory at the University of Calgary, and the transmission link was a 10~m polarization maintaining fiber that did not require any polarization alignment and stabilization. In the second set of experiments, Bob was placed in a laboratory at the Southern Alberta Institute of Technology (SAIT), 3.3~km from the U.~of~Calgary. The link was a 12.4~km underground single mode fiber featuring total loss of 7.3~dB. Polarization stabilization was performed using quantum frames and a fully automatized polarization stabilizer (PS) \cite{LCM+09}. Specifically, each 10~s period was divided into a 0.4~s period used to send a 0.25~s reference light pulse with the 1536.47~nm laser diode $LD_{p}$ (see Fig.~\ref{fig:source}), and a 9.6~s period during which entangled photons were sent. The reference pulse and time-bin qubits were orthogonally polarized and are combined into the link using a polarizing beamsplitter. At Bob's, the reference pulse passes through the polarization stabilizer PS (PSY-101 from General Photonics) and is then reflected through a PBS towards the synchronization detector $D_{s}$. The signal from $D_{s}$ is used to trigger the computer that then enables PS during 50~ms. The pulse is used as a reference to adjust the polarization transformation of PS to maximize the transmission of the single photons through the PBS and through Bob's UTBA. This stabilization system ran uninterrupted throughout an entire day and did not require resetting. 

\begin{figure}[!t]
\includegraphics{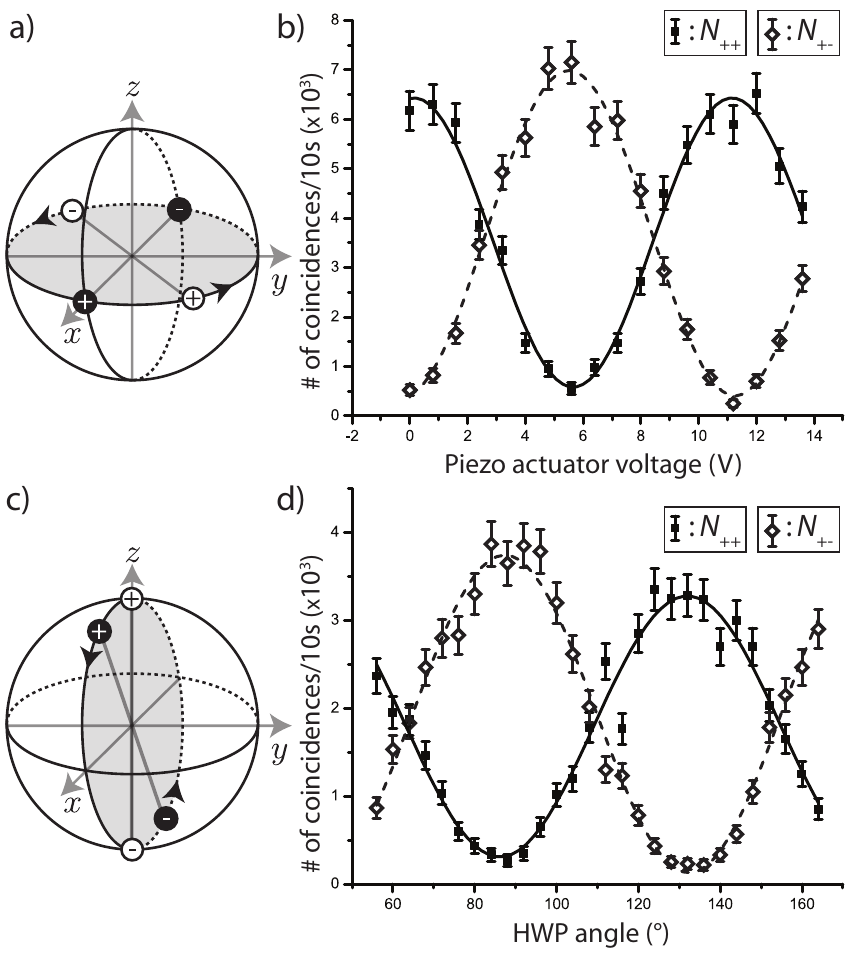}
\caption{Entanglement visibility measurements. a) Bloch sphere depicting Alice's projection measurements (black solid circles with $+$ and $-$ symbols defining Alice's measurement basis) and Bob's measurements (white circles with $+$ and $-$ symbols defining Bob's measurement basis). Bob's measurements are scanned around the equator. b) Subset of the entanglement visibility curves with the settings of a) and with Bob at SAIT. The $y$-axis shows the number of coincidence detections, $N_{++}$ and $N_{+-}$, per 10 seconds, where $N_{++}$ corresponds to coincidences projecting on the state identified by the $+$ black solid circle at Alice and on the $+$ white circle at Bob's, etc. $N_{-+}$ and $N_{--}$ are similar (not shown).
c) Bloch sphere showing Alice's measurement settings (black solid circles) and Bob's measurements (white circles). Alice's measurements are scanned around a great circle orthogonal to the equator. d) Subset of the entanglement visibility curves with the settings of c) and with Bob at SAIT. } 
\label{fig:visibilities} 
\end{figure}

To reveal the presence of entanglement, we first measured the entanglement visibility using projections that are depicted on great circles around the Bloch sphere. First, we positioned Alice's (Bob's) UTBA to measure in the basis $\ket{\pm}_{\phi_{A}} = \frac{1}{\sqrt{2}}(\ket{e} \pm \e^{i\phi_{A}}\ket{\ell})$ ($\ket{\pm}_{\phi_{B}} = \frac{1}{\sqrt{2}}(\ket{e} \pm \e^{i\phi_{B}} \ket{\ell})$). Then, the phase of Bob's UTBA was scanned by varying the voltage applied to the piezo actuator of Bob's UTBA, as illustrated in Fig.~\ref{fig:visibilities}-a. The average visibilities obtained are $91.0\pm2.9$\% with Bob beside Alice and $85.4\pm3.3$\% with Bob at SAIT (Fig.~\ref{fig:visibilities}-b). The decrease in the latter visibility is entirely due to a decreased signal to noise ratio (i.e. photon loss during transmission). This measurement was used to calibrate the phase variation as a function of the voltage applied to the piezo actuator. We also used this measurement to assess the phase stability of the setup. We observed that, typically, the phase did not drift more than $\pi/10$ over 10 minutes, which we set to be the time limit for stability. Then, we positioned Bob's UTBA to project onto $\ket{e}$ and $\ket{\ell}$, and scanned Alice's UTBA to project onto bases aligned along the $xz$ great circle of the Bloch sphere (Fig.~\ref{fig:visibilities}-c). The average visibilities obtained are $95.6\pm1.9$\% for Bob beside Alice and $88.4\pm3.2$\% for Bob at SAIT (Fig.~\ref{fig:visibilities}-d). We point out that this type of measurement is not accessible using a single ``standard'' time-bin qubit analyzer consisting of an unbalanced interferometer and nonpolarizing beamsplitters with fixed splitting ratios. 

\begin{figure}[!t]
\includegraphics{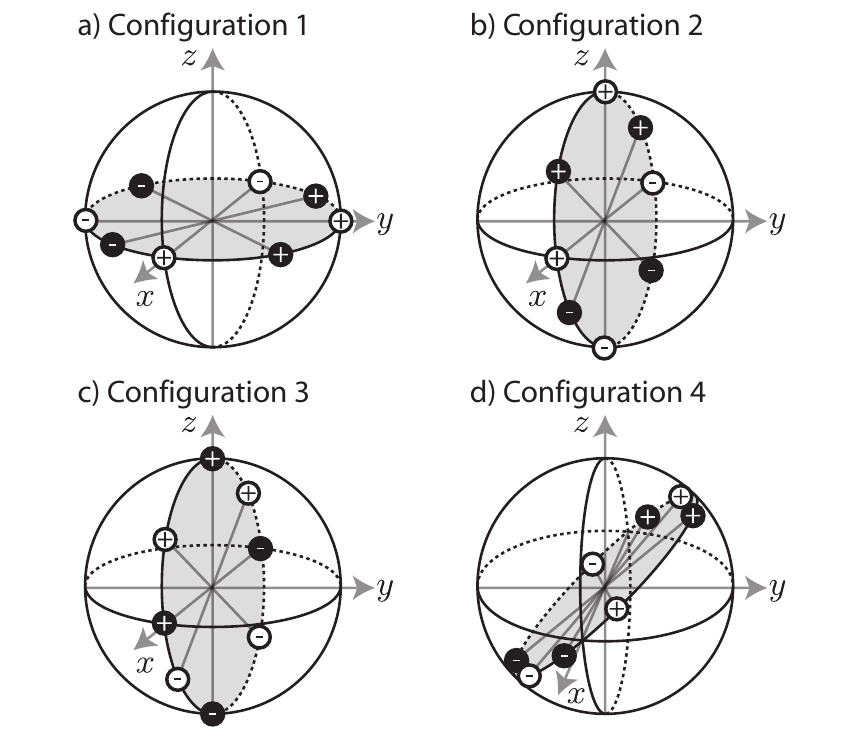}
\caption{The four configurations used to violate the CHSH inequality. Alice's (Bob's) bases are represented by black solid circles (white circles). } \label{fig:CHSH} 
\end{figure}

Next, we used our source to violate the CHSH inequality~\cite{CHSH69} using four configurations that demonstrate the universality of the \mbox{UTBAs}. For each configuration, Alice and Bob projected their qubits randomly onto one of two, configuration specific bases. Each measurement result was registered as either $+1$ or $-1$, and only coincidence detections, one detection at Alice's and one detection at Bob's, were considered. Alice and Bob repeated each experiment with many photons to collect sufficient statistics. We shall use the indices $i$ and $j$, respectively, with $i,j \in \{1,2\}$, to identify Alice's and Bob's measurement basis. Strictly speaking, the basis selection and the measurements should be space-like separated events. Moreover, the overall detection efficiency of Alice and Bob's transmission channels should be above a certain threshold~\cite{Pearle70,Eberhard93}. These conditions were not met in our experiment, and, as is the case in all Bell-inequality tests to date, the violations we obtained are not loophole-free. 

The CHSH inequality reads
\begin{equation} 
S_{CHSH} = |E_{11}+ E_{12} + E_{21} - E_{22}| \le 2, 
\end{equation}
where the correlation coefficient $E_{ij}$ is given by
\begin{equation} 
E_{ij} = \frac{N_{++}^{ij} + N_{--}^{ij} - N_{+-}^{ij} - N_{-+}^{ij} }{N_{++}^{ij} + N_{--}^{ij} + N_{+-}^{ij} + N_{--}^{ij}}.
\end{equation}
$N_{++}^{ij}$ denotes the number of times a coincidence detection with outcome $++$ in the bases $i$ and $j$, respectively, occurred during a fixed time period, etc. Assuming a local variable model, $S$ has an upper bound of two. Quantum mechanics, however, predicts that entangled qubits with entanglement visibility $V>71\%$ can violate this bound: $S_{QM} = 2\sqrt{2}V$. This value is obtained when Alice's and Bob's basis are symmetrically positioned on a great circle around the Bloch sphere as shown in Fig.~\ref{fig:CHSH}. A violation of the CHSH-Bell inequality is obtained for $S > 2$.

The four configurations used are depicted in Fig.~\ref{fig:CHSH}. Alice's (Bob's) measurement bases are shown as black solid circles (white circles) on the Bloch sphere. Previous experiments with time-bin entangled qubits always reported a violation of the CHSH inequality using configuration~1, as in~\cite{TBZG00}. Configurations 2, 3 and 4 constitute the first demonstration of nonlocality with time-bin entangled qubits measured in bases that do not lie exclusively on the equator. Configuration~4 corresponds to configuration~2 after a rotation of $-\pi/8$ around the $y$-axis, followed by a rotation of $-\pi/4$ around the $x$-axis. To the best of our knowledge, this is the first reported violation with such bases for any kind of qubit encoding.

\begin{table}[!t] 
\caption{\label{table:results} Results of the CHSH inequality violations with the four different configurations shown in Fig.~\ref{fig:CHSH}. Columns $S_{A}$ and $S_{S}$ show the values obtained with Bob beside Alice, and Bob at SAIT, respectively. Columns $\times \sigma_{A}$ and $\times\sigma_{S}$ show by how many standard deviations the inequality was violated. Each measurement of $S_{A}$ ($S_{S}$) lasted 160~s (480~s).}
\begin{ruledtabular}
\begin{tabular}{ccccc}
Conf. & $S_{A}\pm\sigma_{A}$ & ($\times \sigma_{A}$) & $S_{S}\pm\sigma_{S}$ & ($\times \sigma_{S}$) \\
\hline
1 & $2.65\pm0.09$ & 7.7 & $2.44\pm0.15$ & 2.9 \\
2 & $2.60\pm0.08$ & 7.5 & $2.40\pm0.15$ & 2.7 \\
3 & $2.65\pm0.09$ & 7.5 & $2.39\pm0.15$ & 2.6 \\
4 & $2.60\pm0.10$ & 6 & $2.39\pm0.15$ & 2.7 \\
\end{tabular}
\end{ruledtabular}
\end{table}

In order to perform the measurements, one could set the phase of each UTBA individually using a frequency stabilized reference laser. However, this is not required as one can set the phase of Bob's UTBA using the results from coincidence measurements. To see this, let us consider the quantum state in the middle time slot shared by Alice and Bob right after PBS2 and PBS5 of Alice's and Bob's UTBA (see Fig.~\ref{fig:UTBA}-b). This state can be written as $\ket{\psi'} = \frac{1}{\sqrt{2}}(\ket{HH}+\e^{i(\phi_{A}+\phi_{B})}\ket{VV})$, where $\phi_{A}$ and $\phi_{B}$ are the relative phases picked up in the two interferometers. We see immediately that setting $\phi_{B}$ to $-\phi_{A}$ yields the desired state $\ket{\psi}$ of Eq.~(\ref{eqn:time-bin}). Accordingly, we varied $\phi_{B}$ using the piezo actuator until we obtained a value of $E_{11}$ that was consistent with the $S$-parameter expected from the measurements of the entanglement visibilities reported above. Then, we proceeded with the measurements of the remaining correlation coefficients. 

Taking into account the measured entanglement visibilities, we expected values of $S$ between $2.57$ and $2.70$ with Bob beside Alice, and between $2.40$ and $2.49$ with Bob at SAIT. As shown in Table~\ref{table:results}, the measured values are in very good agreement with the predictions. Note that the violations (in terms of the number of standard deviations) are limited by the measurement time, which was set to 10~minutes or less to ensure stability of the system.

In conclusion, the results presented here show that entanglement is a concept that is independent of the encoding used, and that it persists when different encodings are converted into each other. This work also opens new possibilities for using time-bin entanglement to perform more stringent test of nonlocality, and to implement new quantum communication protocols that require measurement bases exploring all dimensions of the Bloch sphere.  

We acknowledge SAIT for providing the laboratory, and Vladimir Kiselyov for technical support. This work was supported by NSERC, iCORE, SAIT, GDC, CFI, AAET, QuantumWorks, NATEQ, AIF and CIPI.

\end{document}